# The Relativistic Dynamics of Photon Gas

*Two dynamic equations are established in photon gas, which is just the carrier or medium of electromagnetic waves.*

JiKang Chen †

College of Physics, Nanjing Normal University, Nanjing, 210097, China
† To whom correspondence should be addressed. E-mail: chenjikang@njnu.edu.cn

**Abstract:**
The carrier or medium of electromagnetic waves has been vainly searched for many years, and now it has been caught after the establishment of the dynamic equations in photon gas. The photon's rest mass has been estimated from the cosmic background temperature in space where the photon gas is at an open state of thermal equilibrium, and the photon's proper magnetic moment is calculated from the dynamic equations of photon gas too. As the carrier of electromagnetic waves, the photon gas is a discrete medium at very high frequency, and then the Bohr's electron is hardly to emit energy in wave form and can be stably rounding the nuclei in discrete orbits at lower temperature.

**Introduction:**
Many people have been searching for the carrier or medium of the Maxwell's electromagnetic waves since the last century, and some scientists thought that the photon gas may be the medium, but the dynamic equations hasn't been caught. After the experiment of Michelson and Morley, the searching job was abandoned in the face of the almost universal adherence of physicists to the purely interpretation that the electromagnetic wave is carried by the energy of itself. To prove that the photon gas is the carrier or medium of electromagnetic waves, two dynamic equations must be established, and then some general concepts should be rechecked carefully.

Defined the force based on the acceleration, according to the principle of Einstein's special relativity and between the visual system $S$ where the observer is rest and the proper system $S'$ where the object is rest, the transformations of the acceleration, the inertial mass, the force and the torques (fig. S1-3) can be obtained. When the velocities of objects are comparable with light speed, the Newton's third law should be described as: The interactions (proper forces and proper torques) between two objects are equal and in opposite direction, but the visual ones are unequal in general.

**The estimation about the rest mass of photon**
In the last century, the cosmic microwave background radiation (*1*, *2*) was discovered and the cosmic background temperature ($T_{CB} = 2.725$K) is calculated by the data of COBE satellite (*3*). Supposing every photon has the same rest (stand by) mass $m_s$ and the rest frequency $\nu_s$, when it is moving with the group velocity $c_g$, the parameters of the photon can be described as

$$h\nu_s = m_s c^2, \quad \nu = \gamma \nu_s, \quad c_g^2 = c^2(1 - \gamma^{-2}). \tag{1}$$

In the thermal radiation of a black body, the intensity is composed of two parts: one is the shot noise and the other is the wave noise (*4*), and the Planck formula can be separated mathematically into two parts:

$$I_w(f) = \frac{2\pi c^{-2} h f^3 \exp(-hf/kT)}{\exp(hf/kT) - 1}, \tag{2}$$

$$I_p(\nu) = 2\pi c^{-2} h \nu^3 \exp(-h\nu/kT). \tag{3}$$

Then the number density of the background photons at general temperature is

$$n(\nu) = 8\pi c^{-3} C_T \nu^2 \exp(-h\nu/kT). \tag{4}$$

When the temperature goes higher, the factor $C_T$ approaches to unit, and the distribution of photons agrees with Boson distribution at higher frequency but is different from that one at lower frequency as the wave noise has been separated. Essentially, the difference is caused by photon's interaction which is rather strong for photons with the lower frequency and is weakened rapidly by the increase of the photon's frequency.

Considering the Faraday rotation effects, it is reasonable to suppose that every photon has the same proper magnetic moment, then the magnetic property of photon's rest energy will cause negative mechanical pressure in the photon gas, and the photon's kinetic energy will cause the positive one. The total mechanical pressure should be zero at the cosmic background temperature in the space where the photon gas is at an open state of thermal equilibrium. When the local temperature is higher than the cosmic background temperature, the mechanical pressure of the photon gas is positive and the energy will be diverged. However, when the local temperature is lower than the cosmic background temperature, the mechanical pressure of the photon gas is negative and the energy will be converged. At last, there is a stable cosmic background photon gas fulfilled in our universe. Supposing the average kinetic energy of the photons is equal to its rest energy at the cosmic background temperature, thus

$$\int_{\nu_s}^{\infty} h(\nu - 2\nu_s) n(\nu) d\nu = 0, \ T = T_{CB}. \tag{5}$$

To calculate the integration (Eq. 4, 5) and using the numerical method, the characteristic equation and its solution are

$$a_c^3 + a_c^2 - 2a_c - 6 = 0, \ a_c = h\nu_s / kT_{CB} = 1.8454661. \tag{6}$$

Thus the mass and frequency of a rest photon are

$$m_s = 7.725(3) \times 10^{-40} \text{kg}, \ \nu_s = 1.0479(4) \times 10^{11} \text{Hz}. \tag{7}$$

The interaction between photons forms the special frequency distribution and the unique density of temperature, then the photon gas is fluctuated in large scale, and its radiation property is the same as a black body at the equal temperature.

## The magnetoelectric dynamics of photon gas

If the average velocity of the photons is zero and the photon gas is an isotropic medium in visual system, then the photon gas is moving as an anisotropic medium in the proper system. Since the relativistic transformation of the magnetic moment is complicated, so that the magnetic field **B** should be defined based on the line current, then the *interacting magnetic field* is a tensor of second order in proper system (fig. S4-6) and the magnetic inductivity is a tensor of second order in proper system too, but both the *characteristic magnetic field* and the *characteristic magnetic inductivity* are relatively simple in proper system.

A photon is traveling in the $y_0$ direction with the group velocity $c_g$ in the visual system (fig. S7) and the suitable coordinates are selected. To simplify the mathematical analysis, considering one dimension electromagnetic wave $E_y$–$B_z$ in the visual system and the curl of the electromagnetic field can be calculated (eq. S22-25).

The photon's proper magnetic moment can be equivalent to a circle current according to the principle of correspondence,
$$\mathbf{m}'_p = \hat{x}_0 m'_{px_0} + \hat{y}_0 m'_{py_0} + \hat{z}_0 m'_{pz_0}. \tag{8}$$

The proper force acted on the photon by the curl of magnetic field has been analyzed (fig. S8-11, eq. S37, 44) and can be calculated. The visual force acted on the photon has been calculated and the acceleration of the photon is (eq. S38-40),

$$\frac{\partial c_{gx_0}}{\partial t} = \frac{\gamma^{-2} c^2}{h\nu_s} \frac{\partial B_z}{\partial x} m'_{pz_0} \cos\theta_y,$$

$$\frac{\partial c_{gy_0}}{\partial t} = \frac{\gamma^{-5} c^2}{h\nu_s} \frac{\partial B_z}{\partial x} m'_{pz_0} \sin\theta_y, \tag{9}$$

$$\frac{\partial c_{gz_0}}{\partial t} = -\frac{\gamma^{-2} c^2}{h\nu_s} \frac{\partial B_z}{\partial x} (m'_{px_0} \cos\theta_y + \gamma^{-1} m'_{py_0} \sin\theta_y).$$

The characteristic electric moment of a moving photon is (fig.11, eq. S45, 46)
$$\mathbf{p}_e = c^{-2} \gamma^{-2} c_g (\hat{x}_0 m'_{pz_0} - \hat{z}_0 m'_{px_0}). \tag{10}$$

The partial derivatives of the characteristic electric moment of a moving photon with respect to photon's velocity are (fig. S12, 13, eq. S47-51)

$$\frac{\partial \mathbf{p}_e}{\partial c_{gx_0}} = c^{-2} \gamma^{-2} (\hat{z}_0 m'_{py_0} - \hat{y}_0 m'_{pz_0}),$$

$$\frac{\partial \mathbf{p}_e}{\partial c_{gy_0}} = c^{-2}(3\gamma^{-2} - 2)(\hat{x}_0 m'_{pz_0} - \hat{z}_0 m'_{px_0}), \tag{11}$$

$$\frac{\partial \mathbf{p}_e}{\partial c_{gz_0}} = c^{-2} \gamma^{-2} (\hat{y}_0 m'_{px_0} - \hat{x}_0 m'_{py_0}).$$

The partial derivative of the characteristic electric moment with respect to time can be calculated and averaged by the random properties of the photons in the state of thermal equilibrium (Eq. 9, 11, eq. S52-55), then

$$\overline{\frac{\partial \mathbf{p}_e}{\partial t}} = -\hat{y} \frac{2 m'^2_p}{9 h \nu_s} \frac{\partial B_z}{\partial x} (\gamma^{-4} - \gamma^{-5} + 3\gamma^{-7}). \tag{12}$$

The electric field is the integration of the characteristic electric moment of the photons, and then the partial derivative of the electric field is

$$\frac{\partial \mathbf{D}}{\partial t} = -\int_1^\infty \overline{\frac{\partial \mathbf{p}_e}{\partial t}} n(\gamma) d\gamma. \tag{13}$$

To calculate the integration (Eq. 4) and defined function:
$$K_\varepsilon = \frac{4}{5} \int_1^\infty (\gamma^{-2} - \gamma^{-3} + 3\gamma^{-5}) \exp\frac{-\gamma h \nu_s}{kT} d\gamma. \tag{14}$$

Then the magnetoelectric dynamic equation of the photon gas is
$$\frac{\partial D_y}{\partial t} = \frac{20\pi c^{-3} \nu_s^2 m'^2_p}{9h} C_T K_\varepsilon \frac{\partial B_z}{\partial x}. \tag{15}$$

The change of the electric field is counteractive compared with that in the Maxwell's equation, because the curl of magnetic field is active at the magnetodynamics of photon gas and is passive in the Maxwell's equation that is just the mechanism of Lenz law. At the high temperature and compared with the Maxwell's equation, the magnetic moment of a rest photon is (Eq. 7)

$$C_T K_\varepsilon \to 1, \quad m_p'^2 = \frac{9hc^3}{20\pi v_s^2 \mu_0}, \quad m_p' = 4.305(2) \times 10^{-13} \, \text{Am}^2. \tag{16}$$

Then the simplified magnetoelectric dynamic equation of photon gas at general temperature is

$$\frac{\partial D_y}{\partial t} = \varepsilon_0 c^2 C_T K_\varepsilon \frac{\partial B_z}{\partial x}. \tag{17}$$

## The electromagnetic dynamics of photon gas

A moving photon has two transverse components of characteristic electric moment and the longitudinal component is zero, since the transverse interaction is enlarged by one factor of relativity, so that the *interacting electric moment* is a tensor of second order. The first part of the proper force acted on the photon is the cross production of the interacting electric moment and the curl of electric field, and then the first part of the visual force is obtained (eq. S56-60).

To transform the visual electric field into the characteristic magnetic field in the proper system and calculating its curl this acted on the longitudinal proper magnetic moment of the photon (eq. S26, 27), then the second part of the visual force is obtained and the visual acceleration of the photon acted by the electric field is (eq. S61-65)

$$\frac{\partial c_{gx_0}}{\partial t} = \frac{\gamma^{-3} c_g}{h v_s} \frac{\partial E_y}{\partial x} m'_{pgx_0} \sin\phi_y \sin\theta_y,$$

$$\frac{\partial c_{gy_0}}{\partial t} = \frac{\gamma^{-5} c_g}{h v_s} \frac{\partial E_y}{\partial x} (-m'_{pz_0} \cos\phi_y - m'_{px_0} \sin\phi_y \cos\theta_y), \tag{18}$$

$$\frac{\partial c_{gz_0}}{\partial t} = \frac{\gamma^{-3} c_g}{h v_s} \frac{\partial E_y}{\partial x} (m'_{pz_0} \sin\phi_y \sin\theta_y + m'_{py_0} \cos\phi_y \sin^2\theta_y).$$

The characteristic magnetic moment of a moving photon in the visual system is (fig.11, eq. S41-43)

$$\mathbf{m}_p = \hat{x}_0 \gamma^{-2} m'_{px_0} + \hat{y}_0 \gamma^{-1} m'_{py_0} + \hat{z}_0 \gamma^{-2} m'_{pz_0}. \tag{19}$$

The derivatives of the characteristic magnetic moment of a moving photon with respect to the photon's velocity are (fig. S14, eq. S66-70)

$$\frac{\partial \mathbf{m}_p}{\partial c_{gx_0}} = c_g^{-1}(\gamma^{-1} - \gamma^{-2})(\hat{y}_0 m'_{px_0} + \hat{x}_0 m'_{py_0}),$$

$$\frac{\partial \mathbf{m}_p}{\partial c_{gx_0}} = c_g c^{-2}(-\hat{x}_0 2m'_{px_0} - \hat{y}_0 \gamma m'_{py_0} - \hat{z}_0 2m'_{pz_0}), \tag{20}$$

$$\frac{\partial \mathbf{m}_p}{\partial c_{gx_0}} = c_g^{-1}(\gamma^{-1} - \gamma^{-2})(\hat{z}_0 m'_{py_0} + \hat{y}_0 m'_{pz_0}).$$

The partial derivative of the characteristic magnetic moment with respect to the time can be calculated and averaged by the random properties of photons in the state of thermal equilibrium (Eq. 18, 20, eq. S71, 72), then

$$\overline{\frac{\partial \mathbf{m}_p}{\partial t}} = \hat{z} \frac{m_p'^2}{9 h v_s} \frac{\partial E_y}{\partial x} (3\gamma^{-4} + \gamma^{-5} - 4\gamma^{-7}). \tag{21}$$

The magnetic field is the integration of the characteristic magnetic moment of photons, and then the partial derivative of the magnetic field is

$$\frac{\partial \mathbf{H}}{\partial t} = \int_1^\infty \overline{\frac{\partial \mathbf{m}_p}{\partial t}} n(\gamma) d\gamma . \tag{22}$$

To calculate the integration (Eq. 4) and defined function:

$$K_\mu = \frac{2}{5} \int_1^\infty (3\gamma^{-2} + \gamma^{-3} - 4\gamma^{-5}) \exp\frac{-\gamma h v_s}{kT} d\gamma . \tag{23}$$

Then the electromagnetic dynamic equation of photon gas is

$$\frac{\partial H_z}{\partial t} = \frac{20\pi c^{-3} v_s^2 m_p'^2}{9h} C_T K_\mu \frac{\partial E_y}{\partial x} . \tag{24}$$

The change of the magnetic field is counteractive compared with that in the Maxwell's equation, because the curl of electric field is active at the electrodynamics of photon gas and is passive in the Maxwell's equation that is the mechanism of Lenz law too. At high temperature and compared with the Maxwell's equation, the magnetic moment of rest photon has the same value as Eq. 16, and the simplified electromagnetic dynamic equation of the photon gas at general temperature is

$$\frac{\partial H_z}{\partial t} = \mu_0^{-1} C_T K_\mu \frac{\partial E_y}{\partial x} . \tag{25}$$

The electromagnetic wave equation can be obtained by the association of Eq. 17 and 25,

$$\frac{\partial^2 H_z}{\partial t^2} = c^2 C_T^2 K_\varepsilon K_\mu \frac{\partial^2 H_z}{\partial x^2} . \tag{26}$$

If the speed of electromagnetic wave is just equal to the ultimate velocity $c$, in a good approximation, the temperature factor of the number density of photon gas is

$$C_T = 1 / \sqrt{K_\varepsilon K_\mu} . \tag{27}$$

Then the three dimensional wave equation in photon gas is

$$\partial^2 \mathbf{H} / \partial t^2 + c^2 \nabla \times \nabla \times \mathbf{H} = 0 . \tag{28}$$

Since the electromagnetic field is discrete at the photon scale in photon gas, so that the two dynamic equations and the wave equation are only valued when the wavelength is far larger than the average interspace of photons, and then the photon gas can be treated as a continuous medium.

## The stability of Bohr's electron

There are three factors that significantly influence the wave noise radiated by a black body, rewrite Eq. 2 as

$$I_w = \frac{1}{6} \bar{\varepsilon} PHR, \quad \bar{\varepsilon} = \frac{3kT}{2}, x = \frac{hf}{kT}, P = \frac{x}{e^x - 1}, H = 8\pi c^{-2} f^2, R = e^{-x}. \tag{29}$$

Inner the boundary of a black body, the time duration of the inter-collision of particles is not zero, the collision is "harder" at higher temperature and is "softer" at lower temperature, and the spectrum of Rayleigh-Jeans oscillator is flat at low frequency but is in exponential decay at high frequency, this is just the physics meaning of the Planck factor $P$. $H$ is the classical radiation factor, but factor $R$ restricts the radiation at high frequency when the wavelength is comparable with photon's interspace, where the radiation is dominated by the reactance of photons and the radiated energy is deduced as the frequency goes higher or the temperature goes lower.

Although the random property of Rayleigh-Jeans oscillator in thermal radiation and the harmonic property of Bohr's electron as a radiator are different, but both radiations are dominated by the factor $R$ that is based on the discrete property of photon gas, and the equivalent quality factor of the Bohr's electron as an oscillator is (fig. S15, eq. S73-78)

$$Q_n = \frac{6h^3 c^3 \varepsilon_0^3 n^3}{\pi e^6} \exp \frac{e^4 m_e}{4n^3 h^2 \varepsilon_0^2 kT}. \tag{30}$$

Some calculated values of the logarithmic equivalent quality factors of the Bohr's electron of hydrogen are listed (table S1), and it is clear that more than 20 orbits can be stable at cosmic background temperature, while five orbits can be stable at room temperature, only the first orbit can be stable when the temperature is about $10^4$K, and no orbit can be stable for the electron of hydrogen at far higher temperature.

The electron's orbits are disturbed by the environmental photons, and the energy is exchanged between the photon and the electron, which is equilibrated dynamically on the orbits determined by the theory of Louis de Broglie's standing waves, thus the particle property and the wave property of the electron can be uniformed determinately at atom scale. The disturbances of electron's orbits are going larger when the temperature is rising, the outer orbits will be destroyed first, followed by the middle ones, then the lowest orbit, and at last the atom is ionized completely.

**Discussions**

After the dynamic equations are established, it is clear that the photon gas is the carrier or medium of electromagnetic waves and the main actor is the photons of lower velocity. The photon's velocity is a function of its visual frequency, after being separated from the ultimate speed $c$, the Einstein's special relativity is more solid. Since Eq. 27 is not deduced from the first principle and is only a good approximation, so that the speed of electromagnetic wave in photon gas should be a function of temperature too, and there should be a deviation, small but not zero, between the ultimate speed $c$ and the wave speed at the cosmic background temperature. The photon's velocity will approach the ultimate speed when its visual frequency is very higher and the wave speed will approach the ultimate speed when the temperature of the photon gas is going higher too.

The propagation of electromagnetic wave is influenced by the moving of the photon gas, but the photon's traveling is rather independent to the moving of the photon gas in a short distance when its visual frequency is far higher than the rest one, so the null result had been got in the experiment of Michelson and Morley.

When an electron and a photon are nearby, their magnetic moments will entangle each other and that is just the process of Compton Effect, although the out behavior is like a collision. The jumps of Bohr's electron from one orbit to another will be induced by the tangle between the electron and photon too, but the process is complicated. In the laser with high power the photons are dense and their relative velocities are lower, the tangles between photons will form the photon couples or photon chains or photon arrays, and they can be destroyed by the environmental photons. From Eq. (16),

$$c = \frac{9h^3}{20\pi \mu_0 m_p'^2 m_s^2}. \tag{31}$$

The ultimate speed which is the foundation of Einstein's special relativity can be determined by Planck constant, the proper magnetic moment and the proper mass of the photon, all the three constants are associated with the photon's inner structure that haven't been discovered yet.

## Supporting Materials

**Method S1:** Defined the force based on the acceleration.

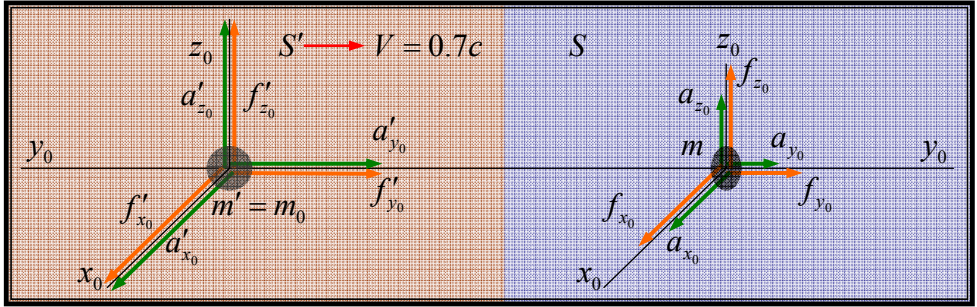

**Figure S1.** Between the visual system $S$ where the observer is rest and the proper system $S'$ where the object is rest, the transformations of the acceleration are

$$a_{y_0} = \gamma^{-3} a'_{y_0}, \quad a_{x_0,z_0} = \gamma^{-2} a'_{x_0,z_0} \quad (S1)$$

The transformation of the inertial mass is $m = \gamma m' = \gamma m_0$ (S2)

Defined the force based on the acceleration, the transformations of the forces are

$$f_{y_0} = \gamma^{-2} f'_{y_0}, \quad f_{x_0,z_0} = \gamma^{-1} f'_{x_0,z_0} \quad (S3)$$

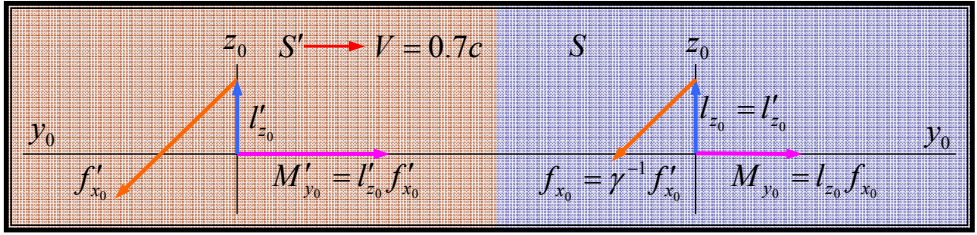

**Figure S2.** The transformation of the longitudinal torque is $M_{y_0} = \gamma^{-1} M'_{y_0}$ (S4)

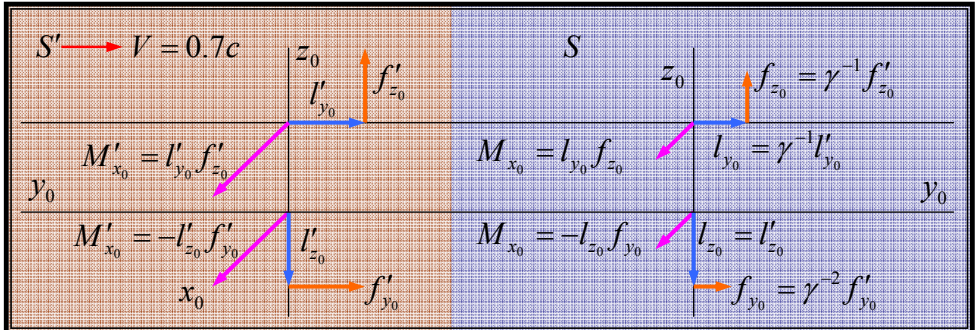

**Figure S3.** The transformations of the transverse torques are $M_{x_0,z_0} = \gamma^{-2} M'_{x_0,z_0}$ (S5)

**Method S2:** The magnetic field **B** should be defined based on the line current.

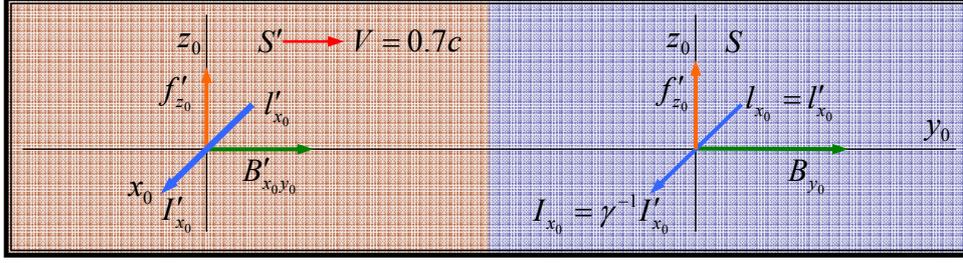

**Figure S4.** To define the magnetic field **B** based on the line current. The relativistic transformations of the line currents are $I_{y_0} = \gamma I'_{y_0}$, $I_{x_0,z_0} = \gamma^{-1} I'_{x_0,z_0}$ (S6).

In the visual system the interaction between the visual transverse current and the longitudinal magnetic field is $f'_{z_0} = I_{x_0} l_{x_0} B_{y_0}$, in the proper system the interaction between the proper transverse current and the longitudinal *interacting magnetic field* is $f'_{z_0} = I'_{x_0} l'_{x_0} B'_{x_0 y_0}$, then the transformations of the longitudinal magnetic field are $B'_{x_0 y_0} = \gamma^{-1} B_{y_0}$ (S7)

and $B'_{z_0 y_0} = \gamma^{-1} B_{y_0}$ (S8)

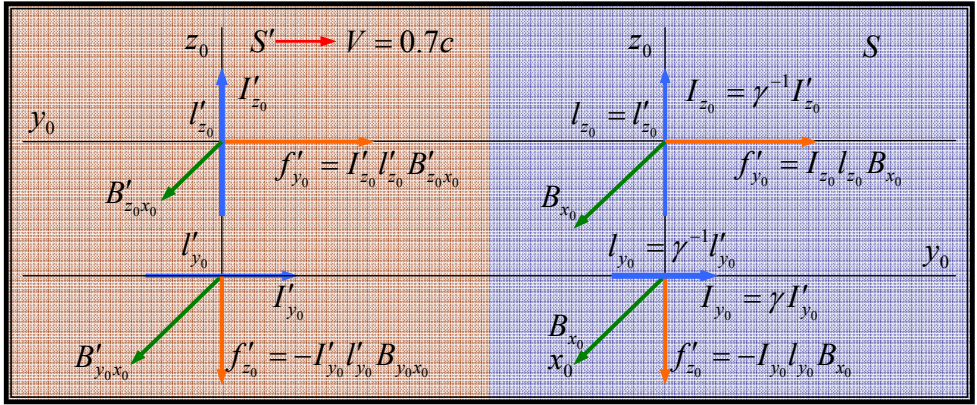

**Figure S5.** On the upper part of the figure the interaction between the visual transverse current and the transverse magnetic field is $f'_{y_0} = I_{z_0} l_{z_0} B_{x_0}$, the interaction between the proper transverse current and the transverse *interacting magnetic field* is $f'_{y_0} = I'_{z_0} l'_{z_0} B'_{z_0 x_0}$, then the transformations of the transverse magnetic field are $B'_{z_0 x_0} = \gamma^{-1} B_{x_0}$ (S9) and $B'_{x_0 z_0} = \gamma^{-1} B_{z_0}$ (S10)

On the lower part of the figure the interaction between the visual longitudinal current and the transverse magnetic field is $f'_{z_0} = -I_{y_0} l_{y_0} B_{x_0}$, the interaction between the proper longitudinal current and the transverse *interacting magnetic field* is $f'_{z_0} = -I'_{y_0} l'_{y_0} B'_{y_0 x_0}$, thus the transformations of the transverse magnetic field are $B'_{y_0 x_0} = B_{x_0}$ (S11) and $B'_{y_0 z_0} = B_{z_0}$ (S12)

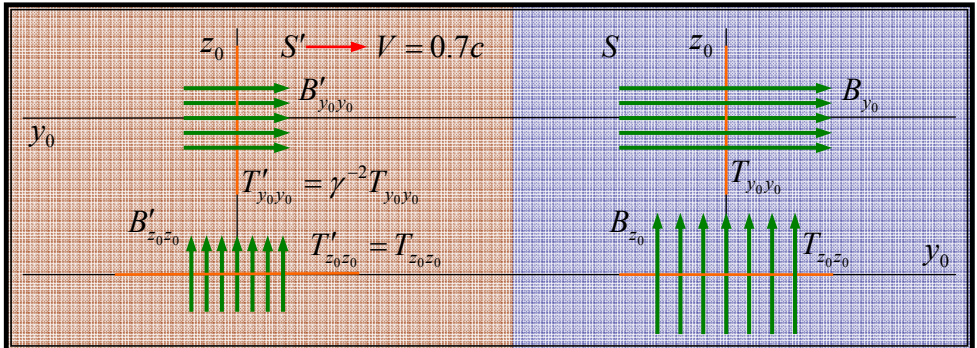

**Figure S6.** To determine the characteristic magnetic field in proper system, consider the magnetic field **B** is a rest object in visual system and is a moving object in proper system, thus the transformations of the forces are the inversion of Eq. S3, and then the transformation of the longitudinal stress is $T'_{y_0 y_0} = \gamma^{-2} T_{y_0 y_0}$, $2T'_{y_0 y_0} = B'_{y_0} H'_{y_0}$, $2T_{y_0 y_0} = B_{y_0} H_{y_0}$, $H'_{y_0} = H_{y_0}$, and the transformation of the longitudinal characteristic magnetic field is $B'_{y_0 y_0} = \gamma^{-2} B_{y_0}$ (S13).

Similarly the transformations of the transverse characteristic magnetic field are $B'_{z_0 z_0} = \gamma^{-1} B_{z_0}$ (S14) and $B'_{x_0 x_0} = \gamma^{-1} B_{x_0}$ (S15). It is clear that the *interacting magnetic field* is a tensor of second order (eq. S7-15) in the proper system:

$$B'_{i_0 j_0} = \begin{vmatrix} B'_{x_0 x_0} & B'_{x_0 y_0} & B'_{x_0 z_0} \\ B'_{y_0 x_0} & B'_{y_0 y_0} & B'_{y_0 z_0} \\ B'_{z_0 x_0} & B'_{z_0 y_0} & B'_{z_0 z_0} \end{vmatrix} = \begin{vmatrix} \gamma^{-1} B_{x_0} & \gamma^{-1} B_{y_0} & \gamma^{-1} B_{z_0} \\ B_{x_0} & \gamma^{-2} B_{y_0} & B_{z_0} \\ \gamma^{-1} B_{x_0} & \gamma^{-1} B_{y_0} & \gamma^{-1} B_{z_0} \end{vmatrix} \quad (S16)$$

The general transformation of the magnetic (potential) field is
$H'_{x_0} = \gamma H_{x_0}$, $H'_{y_0} = H_{y_0}$, $H'_{z_0} = \gamma H_{z_0}$

Then in proper system the magnetic inductivity is a tensor of second order too

$$\mu'_{i_0 j_0} = \frac{B'_{i_0 j_0}}{H'_{j_0}} = \mu_0 \begin{vmatrix} \gamma^{-2} & \gamma^{-1} & \gamma^{-2} \\ \gamma^{-1} & \gamma^{-2} & \gamma^{-1} \\ \gamma^{-2} & \gamma^{-1} & \gamma^{-2} \end{vmatrix} = \gamma^{-2} \mu_0 \begin{vmatrix} 1 & \gamma & 1 \\ \gamma & 1 & \gamma \\ 1 & \gamma & 1 \end{vmatrix} \quad (S17)$$

The characteristic magnetic field in the proper system is
$\mathbf{B}'_c = B'_{i_0} = B'_{i_0 i_0} = \hat{x}_0 \gamma^{-1} B_{x_0} + \hat{y}_0 \gamma^{-2} B_{y_0} + \hat{z}_0 \gamma^{-1} B_{z_0}$ (S18)

The characteristic magnetic inductivity in the proper system is $\mu'_c = \gamma^{-2} \mu_0$ (S19)

**Method S3:** The coordinate transformation for the moving photon.

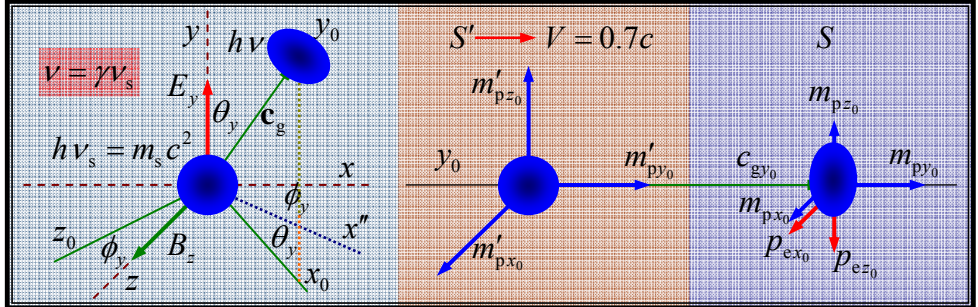

**Figure S7.** When a photon is moving in the $y_0$ direction, the coordinate transformations are

$$\begin{bmatrix} \hat{x} \\ \hat{y} \\ \hat{z} \end{bmatrix} = \begin{bmatrix} \cos\theta_y \cos\phi_y & \sin\theta_y \cos\phi_y & -\sin\phi_y \\ -\sin\theta_y & \cos\theta_y & 0 \\ \cos\theta_y \sin\phi_y & \sin\theta_y \sin\phi_y & \cos\phi_y \end{bmatrix} \begin{bmatrix} \hat{x}_0 \\ \hat{y}_0 \\ \hat{z}_0 \end{bmatrix} \quad (S20)$$

$$\begin{bmatrix} \hat{x}_0 \\ \hat{y}_0 \\ \hat{z}_0 \end{bmatrix} = \begin{bmatrix} \cos\theta_y \cos\phi_y & -\sin\theta_y & \cos\theta_y \sin\phi_y \\ \sin\theta_y \cos\phi_y & \cos\theta_y & \sin\theta_y \sin\phi_y \\ -\sin\phi_y & 0 & \cos\phi_y \end{bmatrix} \begin{bmatrix} \hat{x} \\ \hat{y} \\ \hat{z} \end{bmatrix} \quad (S21)$$

To simplify the mathematical analysis, considering one dimension electromagnetic wave $E_y - B_z$, in the visual system the curl of magnetic field is,
$\text{curl} \mathbf{B} = -\hat{y} \, \partial B_z / \partial x = (\hat{x}_0 \sin\theta_y - \hat{y}_0 \cos\theta_y) \partial B_z / \partial x$ (S22)

In the visual system the curl of electric field is
$\text{curl} \mathbf{E} = \hat{z} \, \partial E_y / \partial x = (\hat{x}_0 \sin\phi_y \cos\theta_y + \hat{y}_0 \sin\phi_y \sin\theta_y + \hat{z}_0 \cos\phi_y) \partial E_y / \partial x$ (S23)

Using Eq. S18, the transformations of the curls of magnetic field are
$\text{curl}'_{x_0} \mathbf{B}'_c = \gamma^{-2} \text{curl}_{x_0} \mathbf{B}$, $\text{curl}'_{y_0} \mathbf{B}'_c = \gamma^{-1} \text{curl}_{y_0} \mathbf{B}$, $\text{curl}'_{z_0} \mathbf{B}'_c = \gamma^{-2} \text{curl}_{z_0} \mathbf{B}$ (S24)

Then in the proper system the curl of magnetic field is (eq. S22, 24),

$$\text{curl}'\mathbf{B}'_c = (\hat{x}_0\gamma^{-2}\sin\theta_y - \hat{y}_0\gamma^{-1}\cos\theta_y)\partial B_z/\partial x \quad (S25)$$

To transform the visual electric field $E_y$ into the characteristic magnetic field in the proper system

$$\mathbf{B}'_{c,E} = \mu'_c\mathbf{H}'_E = \gamma^{-2}\mu_0\mathbf{H}'_E = -\gamma^{-1}c^{-2}\mathbf{c}_g\times\mathbf{E} = -\hat{z}_0\gamma^{-1}c^{-2}c_g\sin\theta_y E_y \quad (S26)$$

For one dimension wave in the $x$ direction (eq. S21)

$$\partial E_y/\partial y = 0, \quad \partial E_y/\partial z = 0,$$

$$\frac{\partial E_y}{\partial x'_0} = \cos\phi_y\cos\theta_y\frac{\partial E_y}{\partial x}, \quad \frac{\partial E_y}{\partial y'_0} = \gamma^{-1}\cos\phi_y\sin\theta_y\frac{\partial E_y}{\partial x}, \quad \frac{\partial E_y}{\partial z'_0} = -\sin\phi_y\frac{\partial E_y}{\partial x}$$

$$\text{curl}'\mathbf{B}'_{c,E} = -c^{-2}c_g(\hat{x}_0\gamma^{-2}\cos\phi_y\sin^2\theta_y - \hat{y}_0\gamma^{-1}\cos\phi_y\sin\theta_y\cos\theta_y)\partial E_y/\partial x \quad (S27)$$

If every photon has a same proper magnetic moment, then in the visual system the moving photon has three components of characteristic magnetic moment and two transverse components of characteristic electric moment.

**Method S4:** The curl of magnetic field interacts with the magnetic moment.

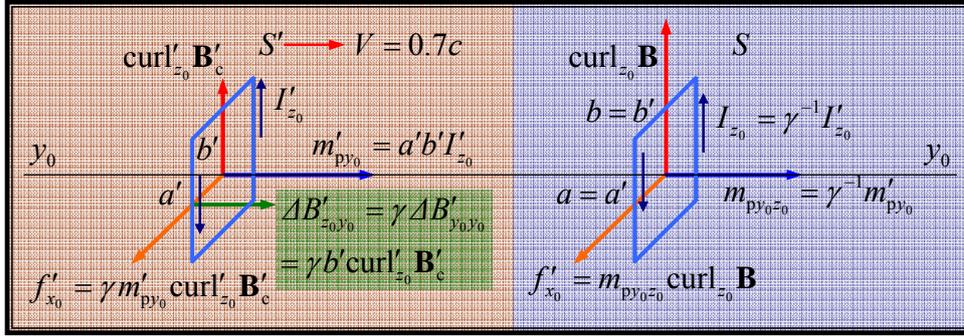

**Figure S8.** The transverse curl of magnetic field interacts with the longitudinal magnetic moment. In the proper system the transverse curl is equivalent to an increment of magnetic field which interacted on one edgy of the longitudinal magnetic moment (eq. S16, 24),

$$\Delta B'_{z_0 y_0} = \gamma\Delta B'_{y_0 y_0} = \gamma^{-1}b'\text{curl}_{z_0}\mathbf{B}_c$$

The proper force acted on the proper magnetic moment is

$$f'_{x_0} = I'_{z_0}a'\Delta B'_{z_0 y_0} = \gamma^{-1}m'_{py_0}\text{curl}_{z_0}\mathbf{B}$$

In the visual system one component of the visual magnetic moment is defined as

$$f'_{x_0} = m_{py_0 z_0}\text{curl}_{z_0}\mathbf{B}, \text{ then } m_{py_0 z_0} = \gamma^{-1}m'_{py_0} \quad (S28), \text{ and } m_{py_0 x_0} = \gamma^{-1}m'_{py_0} \quad (S29)$$

The proper force acted on the longitudinal magnetic moment by the transverse curl of magnetic field can be written in the form of determinant as

$$f'_{B1} = \begin{vmatrix} \hat{x}_0 & C & \hat{z}_0 \\ 0 & \gamma^{-1}m'_{py_0} & 0 \\ \text{curl}_{x_0}\mathbf{B} & 0 & \text{curl}_{z_0}\mathbf{B} \end{vmatrix}_S = \begin{vmatrix} \hat{x}_0\gamma & C & \hat{z}_0\gamma \\ 0 & m'_{py_0} & 0 \\ \text{curl}'_{x_0}\mathbf{B}'_c & 0 & \text{curl}'_{z_0}\mathbf{B}'_c \end{vmatrix}_{S'} \quad (S30)$$

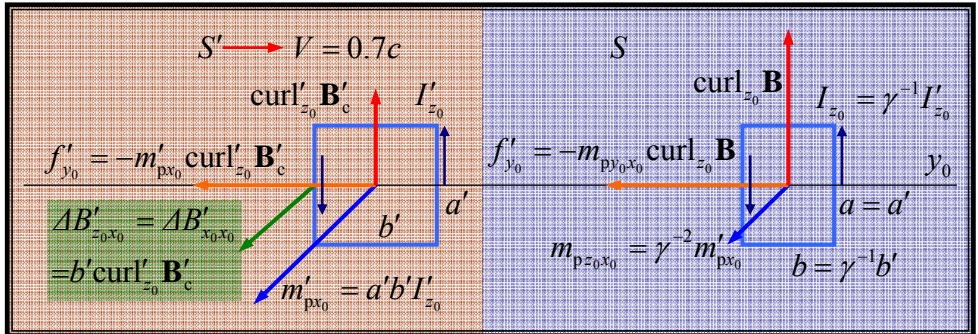

**Figure S9.** The transverse curl of magnetic field interacts with the transverse magnetic moment. In the proper system the transverse curl is equivalent to an increment of magnetic field which interacted on one edgy of the transverse magnetic moment (eq. S16, 24),

$$\Delta B'_{z_0 x_0} = \Delta B'_{x_0 x_0} = \gamma^{-2} b' \text{curl}_{z_0} \mathbf{B}_c$$

The proper force acted on the proper magnetic moment is

$$f'_{y_0} = -I'_{z_0} a' \Delta B'_{z_0 x_0} = -\gamma^{-2} m'_{p x_0} \text{curl}_{z_0} \mathbf{B}$$

In visual system one component of the visual magnetic moment is defined as

$$f'_{y_0} = -m_{p x_0 z_0} \text{curl}_{z_0} \mathbf{B}, \text{ then } m_{p x_0 z_0} = \gamma^{-2} m'_{p x_0} \quad (S31) \text{ and } m_{p z_0 x_0} = \gamma^{-2} m'_{p z_0} \quad (S32)$$

The proper force acted on the transverse magnetic moment by the transverse curl of magnetic field can be written in the form of determinant as

$$f'_{B2} = \begin{vmatrix} C & \hat{y}_0 \gamma^{-2} & C \\ m'_{p x_0} & 0 & m'_{p z_0} \\ \text{curl}_{x_0} \mathbf{B} & 0 & \text{curl}_{z_0} \mathbf{B} \end{vmatrix}_S = \begin{vmatrix} C & \hat{y}_0 & C \\ m'_{p x_0} & 0 & m'_{p z_0} \\ \text{curl}'_{x_0} \mathbf{B}'_c & 0 & \text{curl}'_{z_0} \mathbf{B}'_c \end{vmatrix}_{S'} \quad (S33)$$

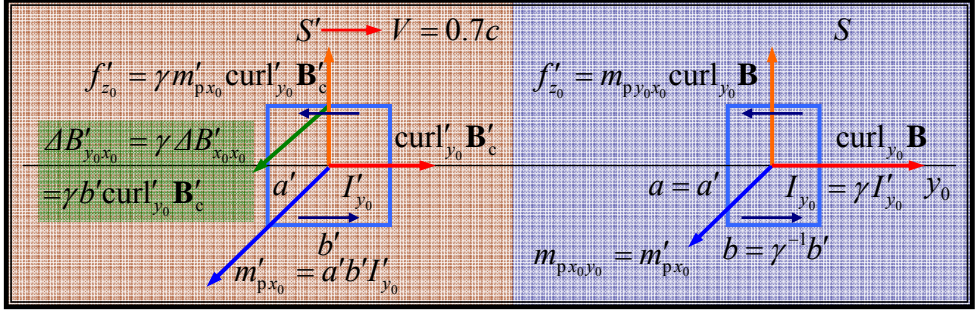

**Figure S10.** The longitudinal curl of magnetic field interacts with the transverse magnetic moment. In proper system the longitudinal curl is equivalent to an increment of magnetic field which interacted on one edgy of the transverse magnetic moment (eq. S16, 24),

$$\Delta B'_{y_0 x_0} = \gamma \Delta B'_{x_0 x_0} = b' \text{curl}_{z_0} \mathbf{B}$$

The proper force acted on the proper magnetic moment is

$$f'_{z_0} = I'_{y_0} a' \Delta B'_{y_0 x_0 c} = m'_{p x_0} \text{curl}_{y_0} \mathbf{B}$$

In visual system one component of the visual magnetic moment is defined as $f'_{z_0} = m_{p x_0 y_0} \text{curl}_{y_0} \mathbf{B}$,

then $m_{p x_0 y_0} = m'_{p x_0}$ (S34), and $m_{p z_0 y_0} = m'_{p z_0}$ (S35)

The proper force acted on the transverse magnetic moment by the longitudinal curl of magnetic field can be written in the form of determinant as

$$f'_{B3} = \begin{vmatrix} \hat{x}_0 & C & \hat{z}_0 \\ m'_{p x_0} & 0 & m'_{p z_0} \\ 0 & \text{curl}_{y_0} \mathbf{B} & 0 \end{vmatrix}_S = \begin{vmatrix} \hat{x}_0 \gamma & C & \hat{z}_0 \gamma \\ m'_{p x_0} & 0 & m'_{p z_0} \\ 0 & \text{curl}'_{y_0} \mathbf{B}'_c & 0 \end{vmatrix}_{S'} \quad (S36)$$

The proper force acted on the photon by the curl of magnetic field (Eq. 8, eq. S30, 33 and 36) is

$$f'_B = \begin{vmatrix} \hat{x}_0 & \hat{y}_0 \gamma^{-2} & \hat{z}_0 \\ m'_{p x_0} & \gamma^{-1} m'_{p y_0} & m'_{p z_0} \\ \text{curl}_{x_0} \mathbf{B} & \text{curl}_{y_0} \mathbf{B} & \text{curl}_{z_0} \mathbf{B} \end{vmatrix}_S = \begin{vmatrix} \hat{x}_0 \gamma & \hat{y}_0 & \hat{z}_0 \gamma \\ m'_{p x_0} & m'_{p y_0} & m'_{p z_0} \\ \text{curl}'_{x_0} \mathbf{B}' & \text{curl}'_{y_0} \mathbf{B}' & \text{curl}'_{z_0} \mathbf{B}' \end{vmatrix}_{S'} \quad (S37)$$

The interaction between the moving photon and the magnetic field is complicated by the tensor property of the *interacting magnetic moment* of the moving photon in the visual system, and is complicated by the tensor property of the *interacting magnetic field* in the proper system where the photon gas is moving as an anisotropic medium.

The visual force acted on the photon (eq. S3, 22 and 37) by the curl of magnetic field is

$$f_B = \begin{vmatrix} \hat{x}_0 \gamma^{-1} & \hat{y}_0 \gamma^{-4} & \hat{z}_0 \gamma^{-1} \\ m'_{p x_0} & \gamma^{-1} m'_{p y_0} & m'_{p z_0} \\ \text{curl}_{x_0} \mathbf{B} & \text{curl}_{y_0} \mathbf{B} & \text{curl}_{z_0} \mathbf{B} \end{vmatrix} = \begin{vmatrix} \hat{x}_0 & \hat{y}_0 \gamma^{-2} & \hat{z}_0 \\ m'_{p x_0} & m'_{p y_0} & m'_{p z_0} \\ \text{curl}'_{x_0} \mathbf{B}' & \text{curl}'_{y_0} \mathbf{B}' & \text{curl}'_{z_0} \mathbf{B}' \end{vmatrix} \quad (S38)$$

$$\frac{f_B}{\partial B_z/\partial x} = \begin{vmatrix} \hat{x}_0\gamma^{-1} & \hat{y}_0\gamma^{-4} & \hat{z}_0\gamma^{-1} \\ m'_{px_0} & \gamma^{-1}m'_{py_0} & m'_{pz_0} \\ \sin\theta_y & -\cos\theta_y & 0 \end{vmatrix} = \begin{pmatrix} \hat{x}_0\gamma^{-1}m'_{pz_0}\cos\theta_y \\ \hat{y}_0\gamma^{-4}m'_{pz_0}\sin\theta_y \\ \hat{z}_0\gamma^{-1}(-m'_{px_0}\cos\theta_y - \gamma^{-1}m'_{py_0}\sin\theta_y) \end{pmatrix} \quad (S39)$$

Then the acceleration of the photon acted by the curl of magnetic field is

$$\begin{bmatrix} \partial c_{gx_0}/\partial t \\ \partial c_{gy_0}/\partial t \\ \partial c_{gz_0}/\partial t \end{bmatrix}_B = \frac{\gamma^{-2}c^2\,\partial B_z/\partial x}{h\nu_s} \begin{bmatrix} m'_{pz_0}\cos\theta_y \\ \gamma^{-3}m'_{pz_0}\sin\theta_y \\ -m'_{px_0}\cos\theta_y - \gamma^{-1}m'_{py_0}\sin\theta_y \end{bmatrix} \quad (S40)$$

**Method S5:** The characteristic moment of a moving photon.

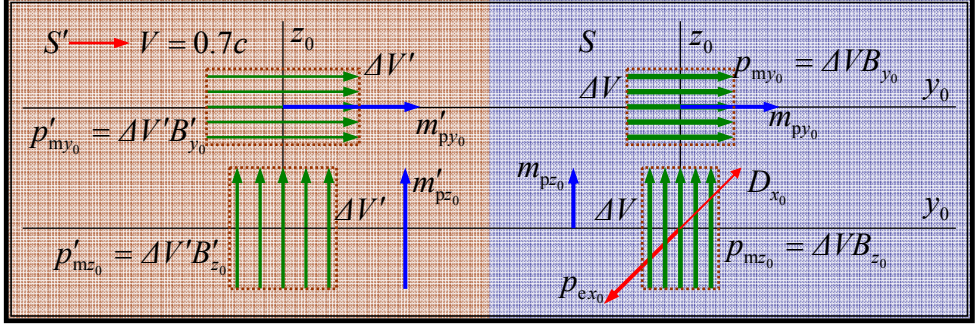

**Figure S11.** The characteristic moments of a moving photon can be deduced from the characteristic magnetic field. On the upper part of the figure, the longitudinal magnetic dipole moment is $p'_{my_0} = \gamma^{-2}\mu_0 m'_{py_0} = \Delta V'B'_{y_0}$ in the proper system, and is $p_{my_0} = \mu_0 m_{py_0} = \Delta V B_{y_0}$ in the visual system. The transformation of the magnetic field is $B_{y_0} = \gamma^2 B'_{y_0}$ (eq. S18), and the transformation of the volume is $\Delta V = \gamma^{-1}\Delta V'$, and then the longitudinal characteristic magnetic moment is $m_{py_0} = \gamma^{-1}m'_{py_0}$ (S41).

On the lower part of the figure, the transverse magnetic dipole moment is $p'_{mz_0} = \gamma^{-2}\mu_0 m'_{pz_0} = \Delta V'B'_{z_0}$ in the proper system, and is $p_{mz_0} = \mu_0 m_{pz_0} = \Delta V B_{z_0}$ in the visual system. The transformations of the magnetic field and the volume are $B_{z_0} = \gamma B'_{z_0}$ (eq. S18) and $\Delta V = \gamma^{-1}\Delta V'$, then the transverse characteristic magnetic moments are

$m_{px_0} = \gamma^{-2}m'_{px_0}, \quad m_{pz_0} = \gamma^{-2}m'_{pz_0}$ (S42)

The *interacting magnetic moment* of a moving photon is a tensor of second order in the visual system (eq. S28, 29, 31, 32, 34, 35, 41 and 42),

$$m_{pi_0 j_0} = \begin{bmatrix} \gamma^{-2}m'_{px_0} & m'_{px_0} & \gamma^{-2}m'_{px_0} \\ \gamma^{-1}m'_{py_0} & \gamma^{-1}m'_{py_0} & \gamma^{-1}m'_{py_0} \\ \gamma^{-2}m'_{pz_0} & m'_{pz_0} & \gamma^{-2}m'_{pz_0} \end{bmatrix} \quad (S43)$$

The proper force acted on the photon by the curl of magnetic field is the cross production of the magnetic moment and the curl of magnetic field in the visual system: (equivalent to eq. 37)

$$\begin{bmatrix} f'_{x_0} \\ f'_{y_0} \\ f'_{z_0} \end{bmatrix} = \begin{bmatrix} m_{py_0z_0}\operatorname{curl}_{z_0}\mathbf{B} - m_{pz_0y_0}\operatorname{curl}_{y_0}\mathbf{B} \\ m_{pz_0x_0}\operatorname{curl}_{x_0}\mathbf{B} - m_{px_0z_0}\operatorname{curl}_{z_0}\mathbf{B} \\ m_{px_0y_0}\operatorname{curl}_{y_0}\mathbf{B} - m_{py_0x_0}\operatorname{curl}_{x_0}\mathbf{B} \end{bmatrix}_S \quad (S44)$$

The moving photon has two characteristic electric moments, illustrated in the lower part of the figure,

$p_{ex_0} = -D_{x_0}\Delta V = \varepsilon_0 c_g p'_{mz_0} = c^{-2}c_g\gamma^{-2}m'_{pz_0}$ (S45) and $p_{ez_0} = -c^{-2}c_g\gamma^{-2}m'_{px_0}$ (S46)

**Method S6:** The derivative of the characteristic electric moment of the photon

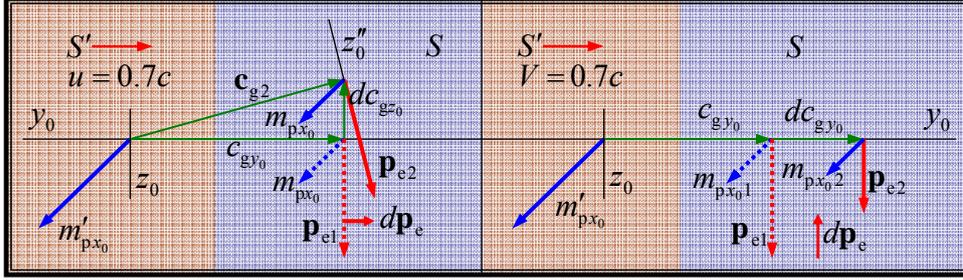

**Figure S12.** The characteristic electric moment of a transverse magnetic moment is changed by the acceleration of the photon.

On the left of the figure (eq. S46) $\mathbf{p}_{e1} = -\hat{z}_0 c^{-2} c_{gy_0} \gamma^{-2} m'_{px_0}$, $\mathbf{p}_{e2} = -\hat{z}''_0 c^{-2} c_{gy_0} \gamma^{-2} m'_{px_0}$, and the derivative of the characteristic electric moment is $d\mathbf{p}_e = \mathbf{p}_{e2} - \mathbf{p}_{e1} = \hat{y}_0 c^{-2} \gamma^{-2} m'_{px_0} dc_{gz_0}$ (S47)

On the right of the figure, the derivative of the characteristic electric moment can be calculated directly (eq. S46), $d\mathbf{p}_e = -\hat{z}_0 c^{-2}(3\gamma^{-2} - 2) m'_{px_0} dc_{gy_0}$ (S48)

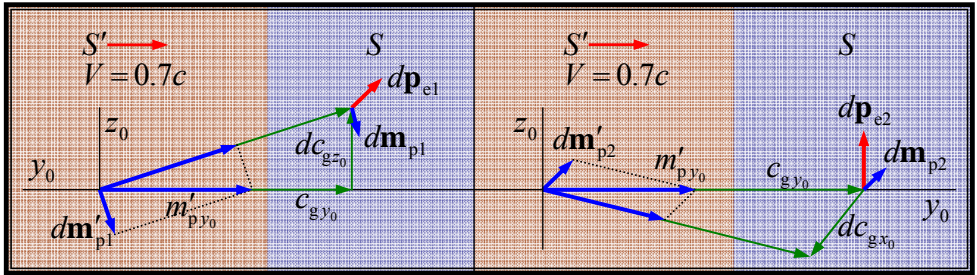

**Figure S13.** The characteristic electric moment of a longitudinal magnetic moment is changed by the acceleration of the photon. On the left of the Figure (Eq. 19),

$d\mathbf{m}'_{p1} = -\hat{z}_0 c_g^{-1} m'_{py_0} dc_{gz_0}$, $d\mathbf{p}_{e1} = c^{-2}\gamma^{-2} \mathbf{c}_g \times d\mathbf{m}'_{p1} = -\hat{x}_0 c^{-2} \gamma^{-2} m'_{py_0} dc_{gz_0}$ (S49)

On the right of the Figure,

$d\mathbf{m}'_{p2} = -\hat{x}_0 c_g^{-1} m'_{py_0} dc_{gx_0}$, $d\mathbf{p}_{e2} = c^{-2}\gamma^{-2} \mathbf{c}_g \times d\mathbf{m}'_{p2} = \hat{z}_0 c^{-2} \gamma^{-2} m'_{py_0} dc_{gx_0}$ (S50)

The partial derivative of the characteristic electric moment with respect to the velocity of the photon is (eq. S47-50)

$$\begin{bmatrix} \partial \mathbf{p}_e / \partial c_{gx_0} \\ \partial \mathbf{p}_e / \partial c_{gy_0} \\ \partial \mathbf{p}_e / \partial c_{gz_0} \end{bmatrix} = c^{-2} \begin{bmatrix} \gamma^{-2}(\hat{z}_0 m'_{py_0} - \hat{y}_0 m'_{pz_0}) \\ (3\gamma^{-2} - 2)(\hat{x}_0 m'_{pz_0} - \hat{z}_0 m'_{px_0}) \\ \gamma^{-2}(\hat{y}_0 m'_{px_0} - \hat{x}_0 m'_{py_0}) \end{bmatrix}$$ (S51)

The total derivative of the characteristic electric moment with respect to time is (eq. S40, 51)

$$\frac{\partial \mathbf{p}_e}{\partial t} = \frac{\gamma^{-4} \partial B_z / \partial x}{h v_s} \begin{bmatrix} +(\hat{z}_0 m'_{py_0} - \hat{y}_0 m'_{pz_0}) m'_{pz_0} \cos\theta_y \\ + \gamma^{-1}(3\gamma^{-2} - 2)(\hat{x}_0 m'_{pz_0} - \hat{z}_0 m'_{px_0}) m'_{pz_0} \sin\theta_y \\ - (\hat{y}_0 m'_{px_0} - \hat{x}_0 m'_{py_0})(m'_{px_0} \cos\theta_y + \gamma^{-1} m'_{py_0} \sin\theta_y) \end{bmatrix}$$ (S52)

In the state of thermal equilibrium the orientation of photon's proper magnetic moment is random, and the average productions of the components of photon's proper magnetic moment are

$$\overline{m'_{pi} m'_{pj}} = \begin{cases} m'^2_p/3 & i = j \\ 0 & i \neq j \end{cases}$$ (S53)

Then the average derivative of the characteristic electric moment with respect to time is

$$\overline{\frac{\partial \mathbf{p}_e}{\partial t}} = \frac{\gamma^{-4} \partial B_z / \partial x}{h v_s} \frac{m'^2_p}{3} \left( \hat{x}_0 \gamma^{-1}(3\gamma^{-2} - 1)\sin\theta_y - \hat{y}_0 2\cos\theta_y \right)$$ (S54)

In the state of thermal equilibrium the directions distribution of the moving photons are random too, to transform the equation into the $(x, y, z)$ coordinate and to average it in spherical coordinates, then the average derivative of the characteristic electric moment of the photons with the same frequency is obtained.

$$\overline{\frac{\partial \mathbf{p}_e}{\partial t}} = -\hat{y}\frac{\gamma^{-4} \partial B_z/\partial x}{h\nu_s} \frac{2m_p'^2}{9}\left(1 - \gamma^{-1} + 3\gamma^{-3}\right) \quad (S55)$$

**Method S7:** The curl of electric field interacts with the electric moment of a photon

A moving photon has two transverse components of the characteristic electric moments (fig. S7, 11, eq. S45, 46) $\quad p_{ex_0} = c^{-2}c_g\gamma^{-2}m'_{pz_0}, \quad p_{ez_0} = -c^{-2}c_g\gamma^{-2}m'_{px_0}$ (S56)

Since the longitudinal component is zero and the transverse interaction is enlarged by one factor of relativity that is just like the situation of moving electron, so that the *interacting electric moment* of a moving photon is a tensor of second order

$$p_{ei_0j_0} = \begin{bmatrix} p_{ex_0x_0} & p_{ex_0y_0} & p_{ex_0z_0} \\ 0 & 0 & 0 \\ p_{ez_0x_0} & p_{ez_0y_0} & p_{ez_0z_0} \end{bmatrix} = \begin{bmatrix} \gamma p_{ex_0} & p_{ex_0} & \gamma p_{ex_0} \\ 0 & 0 & 0 \\ \gamma p_{ez_0} & p_{ez_0} & \gamma p_{ez_0} \end{bmatrix} \quad (S57)$$

The first part of the proper force acted on the photon is the cross production of the electric moment and the curl of electric field and can be written in a clear form of determinant.

$$f'_{E1} = \begin{vmatrix} \hat{x}_0\gamma & \hat{y}_0 & \hat{z}_0\gamma \\ p_{ex_0} & 0 & p_{ez_0} \\ \mathrm{curl}_{x_0}\mathbf{E} & \mathrm{curl}_{y_0}\mathbf{E} & \mathrm{curl}_{z_0}\mathbf{E} \end{vmatrix} \quad (S58)$$

The visual force acted on the photon by the curl of electric field is

$$f_{E1} = \begin{vmatrix} \hat{x}_0 & \hat{y}_0\gamma^{-2} & \hat{z}_0 \\ p_{ex_0} & 0 & p_{ez_0} \\ \mathrm{curl}_{x_0}\mathbf{E} & \mathrm{curl}_{y_0}\mathbf{E} & \mathrm{curl}_{z_0}\mathbf{E} \end{vmatrix} \quad (S59)$$

And it can be calculated (eq. S23),

$$f_{E1} = c^{-2}c_g\gamma^{-2}\frac{\partial E_y}{\partial x}\begin{vmatrix} \hat{x}_0 & \hat{y}_0\gamma^{-2} & \hat{z}_0 \\ m'_{pz_0} & 0 & -m'_{px_0} \\ \sin\phi_y\cos\theta_y & \sin\phi_y\sin\theta_y & \cos\phi_y \end{vmatrix} \quad (S60)$$

To transform the visual electric field into the characteristic magnetic field in proper system (eq. S26), then the interaction between the longitudinal magnetic moment and the curl of magnetic field in proper system is (eq. S38)

$$f'_{E2} = \begin{vmatrix} \hat{x}_0\gamma & \hat{y}_0 & \hat{z}_0\gamma \\ 0 & m'_{py_0} & 0 \\ \mathrm{curl}'_{x_0}\mathbf{B}'_{c,E} & \mathrm{curl}'_{y_0}\mathbf{B}'_{c,E} & \mathrm{curl}'_{z_0}\mathbf{B}'_{c,E} \end{vmatrix} \quad (S61)$$

And it can be calculated (eq. S27),

$$f'_{E2} = \hat{z}_0\gamma^{-1}c^{-2}c_gm'_{py_0}\cos\phi_y\sin^2\theta_y\,\partial E_y/\partial x \quad (S62)$$

The second part of the visual force acted on the photon by the electric field is

$$f_{E2} = \hat{z}_0\gamma^{-2}c^{-2}c_gm'_{py_0}\cos\phi_y\sin^2\theta_y\,\partial E_y/\partial x \quad (S63)$$

The total visual force acted on the photon by the electric field is (eq. S60, 63)

$$f_E = c^{-2}c_g\gamma^{-2}\frac{\partial E_y}{\partial x}\begin{pmatrix} \hat{x}_0 m'_{px_0}\sin\phi_y\sin\theta_y \\ -\hat{y}_0\gamma^{-2}(m'_{px_0}\sin\phi_y\cos\theta_y + m'_{pz_0}\cos\phi_y) \\ \hat{z}_0(m'_{pz_0}\sin\phi_y\sin\theta_y + m'_{py_0}\cos\phi_y\sin^2\theta_y) \end{pmatrix} \quad (S64)$$

Then the visual acceleration of the photon acted by the electric field is

$$\begin{bmatrix} \partial c_{gx_0}/\partial t \\ \partial c_{gy_0}/\partial t \\ \partial c_{gz_0}/\partial t \end{bmatrix}_E = \frac{\gamma^{-3} c_g}{h\nu_s} \frac{\partial E_y}{\partial x} \begin{bmatrix} m'_{px_0} \sin\phi_y \sin\theta_y \\ -\gamma^{-2} m'_{pz_0} \cos\phi_y - \gamma^{-2} m'_{px_0} \sin\phi_y \cos\theta_y \\ m'_{pz_0} \sin\phi_y \sin\theta_y + m'_{py_0} \cos\phi_y \sin^2\theta_y \end{bmatrix} \quad (S65)$$

**Method S8:** The derivative of the characteristic magnetic moment of the photon

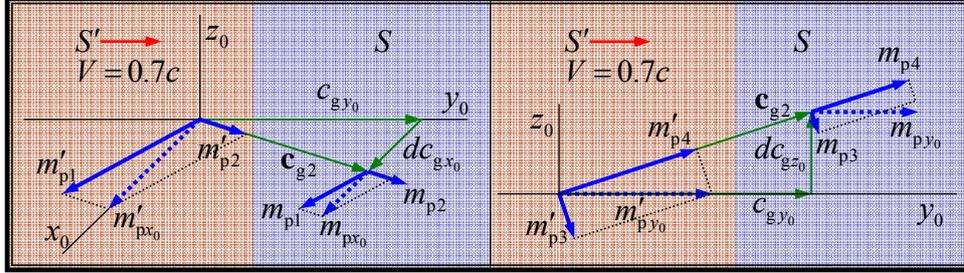

**Figure S14.** The characteristic magnetic moment in visual system is changed by the transverse acceleration of the photon. For the transverse magnetic moment on the left of the figure (Eq. 19),

$d\mathbf{m}_p(m'_{px_0}) = m_{p1} + m_{p2} - m_{px_0} = \gamma^{-2} m'_{p1} + \gamma^{-1} m'_{p2} - \gamma^{-2} m'_{px_0} = \hat{y}_0 c_g^{-1}(\gamma^{-1} - \gamma^{-2}) m'_{px_0} dc_{gx_0}$ (S66)

And then $d\mathbf{m}_p(m'_{pz_0}) = \hat{y}_0 c_g^{-1}(\gamma^{-1} - \gamma^{-2}) m'_{pz_0} dc_{gz_0}$ (S67)

For the longitudinal magnetic moment on the right of the figure (Eq. 19),

$d\mathbf{m}_p(m'_{py_0}) = m_{p3} + m_{p4} - m_{py_0} = \gamma^{-2} m'_{p3} + \gamma^{-1} m'_{p4} - \gamma^{-1} m'_{py_0} = \hat{z}_0 c_g^{-1}(\gamma^{-1} - \gamma^{-2}) m'_{py_0} dc_{gz_0}$ (S68)

And then $d\mathbf{m}_p(m'_{py_0}) = \hat{x}_0 c_g^{-1}(\gamma^{-1} - \gamma^{-2}) m'_{py_0} dc_{gx_0}$ (S69)

The effect of longitudinal acceleration of the photon can be calculated directly.
Then the partial derivatives of the characteristic magnetic moment with respect to the velocity of the photon are (eq. S66-69)

$$\begin{cases} \partial \mathbf{m}_p / \partial c_{gx_0} = \hat{y}_0 c_g^{-1}(\gamma^{-1} - \gamma^{-2}) m'_{px_0} + \hat{x}_0 c_g^{-1}(\gamma^{-1} - \gamma^{-2}) m'_{py_0} \\ \partial \mathbf{m}_p / \partial c_{gy_0} = -\hat{x}_0 2 c_g c^{-2} m'_{px_0} - \hat{y}_0 \gamma c_g c^{-2} m'_{py_0} - \hat{z}_0 2 c_g c^{-2} m'_{pz_0} \\ \partial \mathbf{m}_p / \partial c_{gz_0} = \hat{z}_0 c_g^{-1}(\gamma^{-1} - \gamma^{-2}) m'_{py_0} + \hat{y}_0 c_g^{-1}(\gamma^{-1} - \gamma^{-2}) m'_{pz_0} \end{cases} \quad (S70)$$

The derivative of the characteristic magnetic moment with respect to the time is (eq. S65, 70)

$$\frac{\partial \mathbf{m}_p}{\partial t} = \frac{\gamma^{-3}}{h\nu_s} \frac{\partial E_y}{\partial x} \begin{bmatrix} (\gamma^{-1} - \gamma^{-2})(\hat{y}_0 m'_{px_0} + \hat{x}_0 m'_{py_0}) m'_{px_0} \sin\phi_y \sin\theta_y \\ + \gamma^{-2}\beta^2(\hat{x}_0 2 m'_{px_0} + \hat{y}_0 \gamma m'_{py_0} + \hat{z}_0 2 m'_{pz_0})(m'_{pz_0} \cos\phi_y + m'_{px_0} \sin\phi_y \cos\theta_y) \\ + (\gamma^{-1} - \gamma^{-2})(\hat{z}_0 m'_{py_0} + \hat{y}_0 m'_{pz_0})(m'_{pz_0} \sin\phi_y \sin\theta_y + m'_{py_0} \cos\phi_y \sin^2\theta_y) \end{bmatrix}$$

In the state of thermal equilibrium the orientation of photon's proper magnetic moment is random, and then the average derivative is (eq. S53)

$$\overline{\frac{\partial \mathbf{m}_p}{\partial t}} = \frac{\gamma^{-3}}{h\nu_s} \frac{\partial E_y}{\partial x} \frac{m'^2_p}{3} \begin{bmatrix} (\gamma^{-1} - \gamma^{-2})\hat{y}_0 \sin\phi_y \sin\theta_y \\ 2\gamma^{-2}(1-\gamma^{-2})(\hat{x}_0 \sin\phi_y \cos\theta_y + \hat{z}_0 \cos\phi_y) \\ (\gamma^{-1} - \gamma^{-2})(\hat{z}_0 \cos\phi_y \sin^2\theta_y + \hat{y}_0 \sin\phi_y \sin\theta_y) \end{bmatrix} \quad (S71)$$

In the state of thermal equilibrium the directions distribution of the moving photons are random too, to transform the equation into the (x, y, z) coordinate and to average it in spherical coordinates, then the average derivative of the characteristic magnetic moment of the photons with the same frequency is obtained.

$$\overline{\frac{\partial \mathbf{m}_p}{\partial t}} = \hat{z} \frac{\gamma^{-4} m'^2_p}{9h\nu_s} \frac{\partial E_y}{\partial x}(3 + \gamma^{-1} - 4\gamma^{-3}) \quad (S72)$$

**Method S9:** the Bohr's electron as an oscillator and its equivalent quality factor

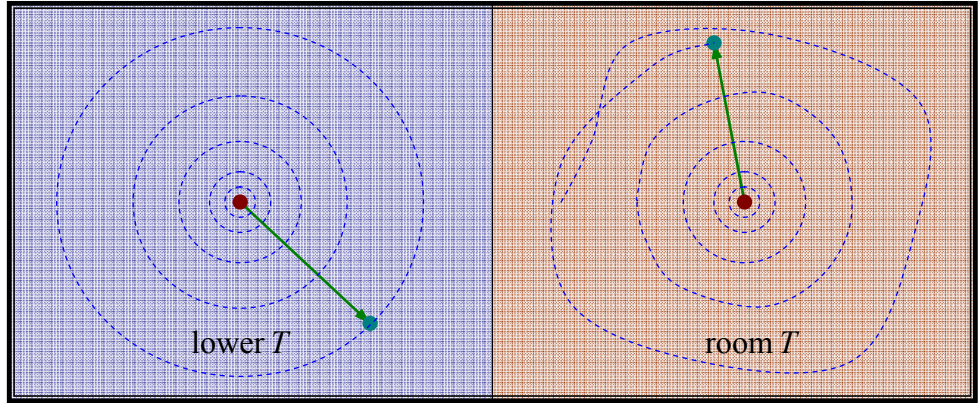

**Figure S15.** The Bohr's electron is moving around the nuclei of hydrogen.

According to the Bohr's theory the orbit radii are $r_n = \dfrac{n^2 h^2}{4\pi^2 m_e} \left( \dfrac{e^2}{4\pi\varepsilon_0} \right)^{-1}$ (73)

The frequencies of the Bohr's electron as an electric dipole are $f_n = \dfrac{4\pi^2 m_e}{n^3 h^3} \left( \dfrac{e^2}{4\pi\varepsilon_0} \right)^2$ (74)

The kinetic energy of the Bohr's electron is $\varepsilon_{k,n} = 2\pi^2 m_e f_n^2 r_n^2$ (75)

Classical radiation of the Bohr's electron is $P'_{w,n} = \dfrac{8\pi^3 \mu_0 e^2 f_n^4}{3c} r_n^2$ (76)

Although the random property of Rayleigh-Jeans oscillator in thermal radiation and the harmonic property of the Bohr's electron as a radiator are different, but both radiations are dominated by the factor $R$ which is based on the discrete property of photon gas and is defined at Eq. 56, then the wave radiation of the Bohr's electron is

$P_{w,n} = P'_{w,n} R = P'_{w,n} \exp \dfrac{-h f_n}{kT}$ (77)

The equivalent quality factor of the Bohr's electron as an oscillator is

$Q_n = \dfrac{2 f_n \varepsilon_{k,n}}{P_{w,n}} = \dfrac{6 h^3 c^3 \varepsilon_0^3 n^3}{\pi e^6} \exp \dfrac{e^4 m_e}{4 n^3 h^2 \varepsilon_0^2 kT}$ (78)

**Table S1**  The logarithmic equivalent quality factors of the Bohr's electron, $\log Q_n$

| n | 3K | 10K | 30K | 100K | 300K | 1000K | 3000K | 10000K |
|---|------|-------|------|------|------|-------|-------|--------|
| 1 | 45721 | 13720 | 4577 | 1377 | 463 | 143 | 51.5 | 19.5 |
| 2 | 5721 | 1721 | 578 | 178 | 63.8 | 23.8 | 12.4 | 8.4 |
| 3 | 1700 | 515 | 177 | 58.0 | 24.1 | 12.3 | 8.9 | 7.7 |
| 4 | 722 | 221 | 79.0 | 29.0 | 14.7 | 9.7 | 8.3 | |
| 5 | 374 | 118 | 44.5 | 18.9 | 11.5 | 9.0 | | |
| 10 | 54.5 | 22.5 | 13.4 | 10.1 | 9.2 | | | |
| 15 | 22.0 | 13.4 | 10.7 | 9.7 | | | | |
| 20 | 15.4 | 11.4 | 10.3 | | | | | |